\documentclass[conference]{IEEEtran}
\usepackage{multicol}
\usepackage{etoolbox}
\makeatletter
\patchcmd{\@makecaption}
  {\scshape}
  {}
  {}
  {}
\makeatletter
\patchcmd{\@makecaption}
  {\\}
  {.\ }
  {}
  {}
\makeatother

\usepackage{amsfonts}
\usepackage{amsmath}
\usepackage{mathrsfs}
\usepackage{mathtools}
\usepackage{amsfonts}
\usepackage{amssymb}
\usepackage{graphicx}
\usepackage{epsfig}
\usepackage{psfrag}{}
\usepackage{array}
\usepackage{cases}
\usepackage{eufrak}
\usepackage{cite,graphicx,amssymb,color}
\usepackage{algorithmic}
\usepackage{algorithm}
\usepackage{setspace}
\usepackage{subfigure}
\usepackage{bm}
\usepackage{multirow}
\usepackage{threeparttable}
\usepackage{array}
\usepackage{makecell}
\usepackage{soul}
\usepackage{float}
\usepackage{booktabs}
\newfloat{figtab}{htb}{fgtb}
\makeatletter
  \newcommand\figcaption{\def\@captype{figure}\caption}
  \newcommand\tabcaption{\def\@captype{table}\caption}
\makeatother

\newtheorem{Thm}{Theorem}

\newtheorem{Exam}{Example}

\newtheorem{Prob}{Problem}
\newtheorem{Rem}{Remark}

\newtheorem{Proof}{Proof}

\IEEEoverridecommandlockouts
\begin{document}
\title{Rate Splitting for General Multicast}

\author{\IEEEauthorblockN{Lingzhi Zhao, Ying Cui}\IEEEauthorblockA{Shanghai Jiao Tong Univ., CN}\and\IEEEauthorblockN{Sheng Yang}\IEEEauthorblockA{Paris-Saclay Univ., FR}\and\IEEEauthorblockN{Shlomo Shamai (Shitz)}\IEEEauthorblockA{Technion-Israel Inst. of Tech., IL}\and\IEEEauthorblockN{Yunbo Han, Yunfei Zhang}\IEEEauthorblockA{Tencent Tech., CN}}


\maketitle

\begin{abstract}
Immersive video, such as virtual reality (VR) and multi-view videos, is growing in popularity. Its wireless streaming is an instance of general multicast, extending conventional unicast and multicast, whose effective design is still open. This paper investigates the \textcolor{black}{optimization} of general rate splitting with linear beamforming for general multicast. Specifically, we consider a multi-carrier single-cell wireless network where a multi-antenna base station (BS) communicates to multiple single-antenna users via general multicast. Linear beamforming is adopted at the BS, and joint decoding is adopted at each user. We consider the maximization of the weighted sum rate, which is a challenging nonconvex problem. Then, we propose an iterative algorithm for the problem to obtain a KKT point using the concave-convex procedure (CCCP). The proposed optimization framework generalizes the existing ones for rate splitting for various types of services. Finally, we numerically show substantial gains of the proposed solutions over existing schemes and reveal the design insights of general rate splitting for general multicast.
\end{abstract}


\section{Introduction}
Conventional mobile Internet services include (traditional) video, audio, web browsing, social networking, software downloading, etc. These services can be supported by unicast, single-group multicast, and multi-group multicast. 
Immersive video, such as 360 video 
and multi-view video 
is growing in popularity.
When watching a tiled 360 video, 
the tiles in a user's current field-of-view (FoV) plus a safe margin are usually transmitted to the user in case of an FoV change. 
On the other hand, when watching a multi-view video, 
a user's current view and adjacent views are usually transmitted to the user in case of a view switch. 
When streaming a popular immersive video 
to multiple users simultaneously, multiple messages (e.g., tiles for 360 video and views for multi-view video) are transmitted to each user, and one message may be intended for multiple users\cite{TWC21,TCOM20}, as illustrated in Fig. 1. \textcolor{black}{This emerging service plays an important role in online gaming, self-driving, and cloud meeting, etc. but cannot perfectly adapt to the conventional transmission schemes mentioned above.}
This motivates us to consider general multicast (also referred to as \textcolor{black}{general connection\cite{TON17}} and general groupcast \cite{ISIT17}) where one message can be intended for any user. Clearly, general multicast includes the three conventional transmission schemes as special cases.

\begin{figure}[t]
\begin{center}
 \subfigure[\small{360 video}]
 {\resizebox{4.3cm}{!}{\includegraphics{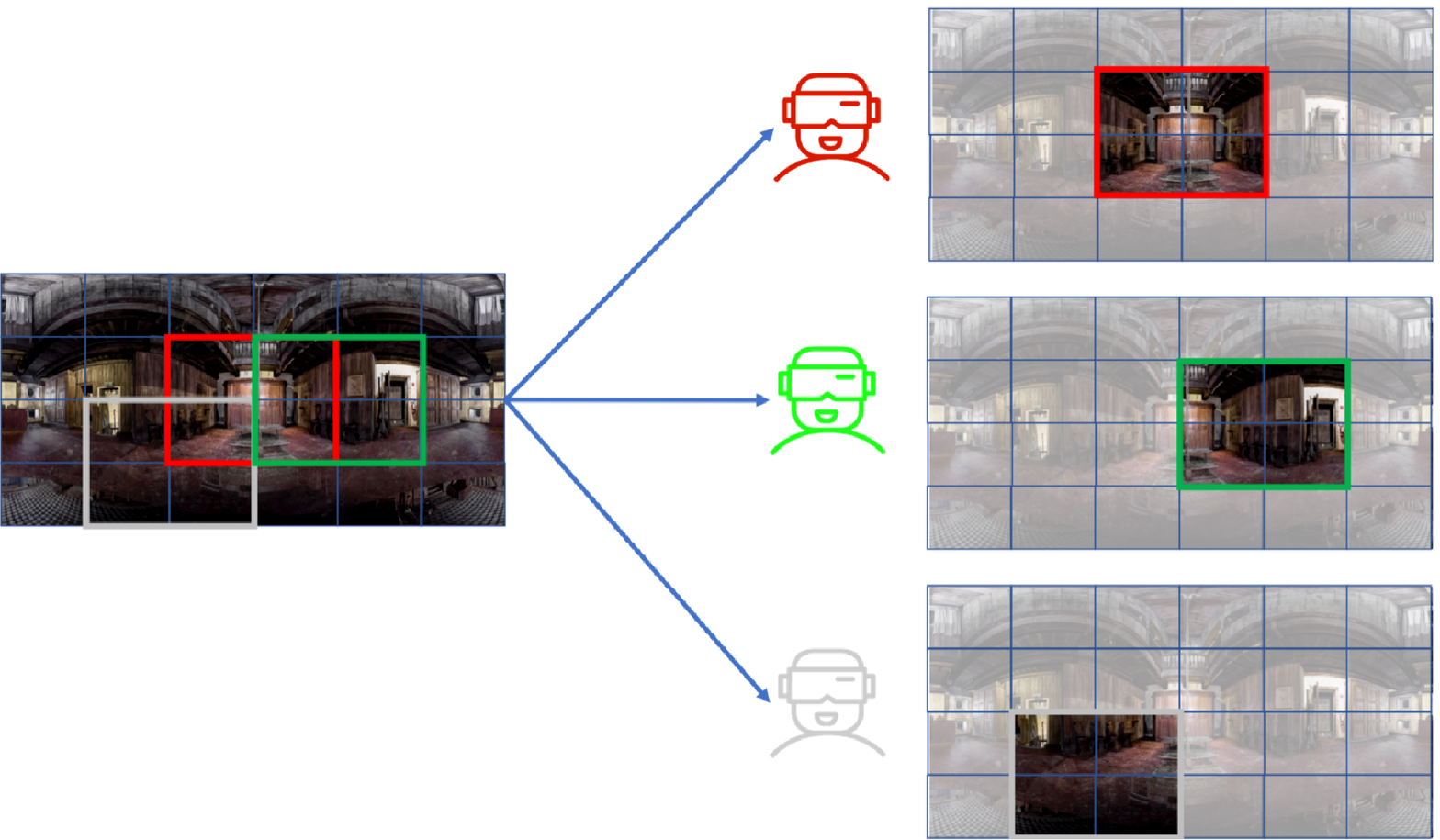}}}
 \subfigure[\small{Multi-view video}]
 {\resizebox{4.3cm}{!}{\includegraphics{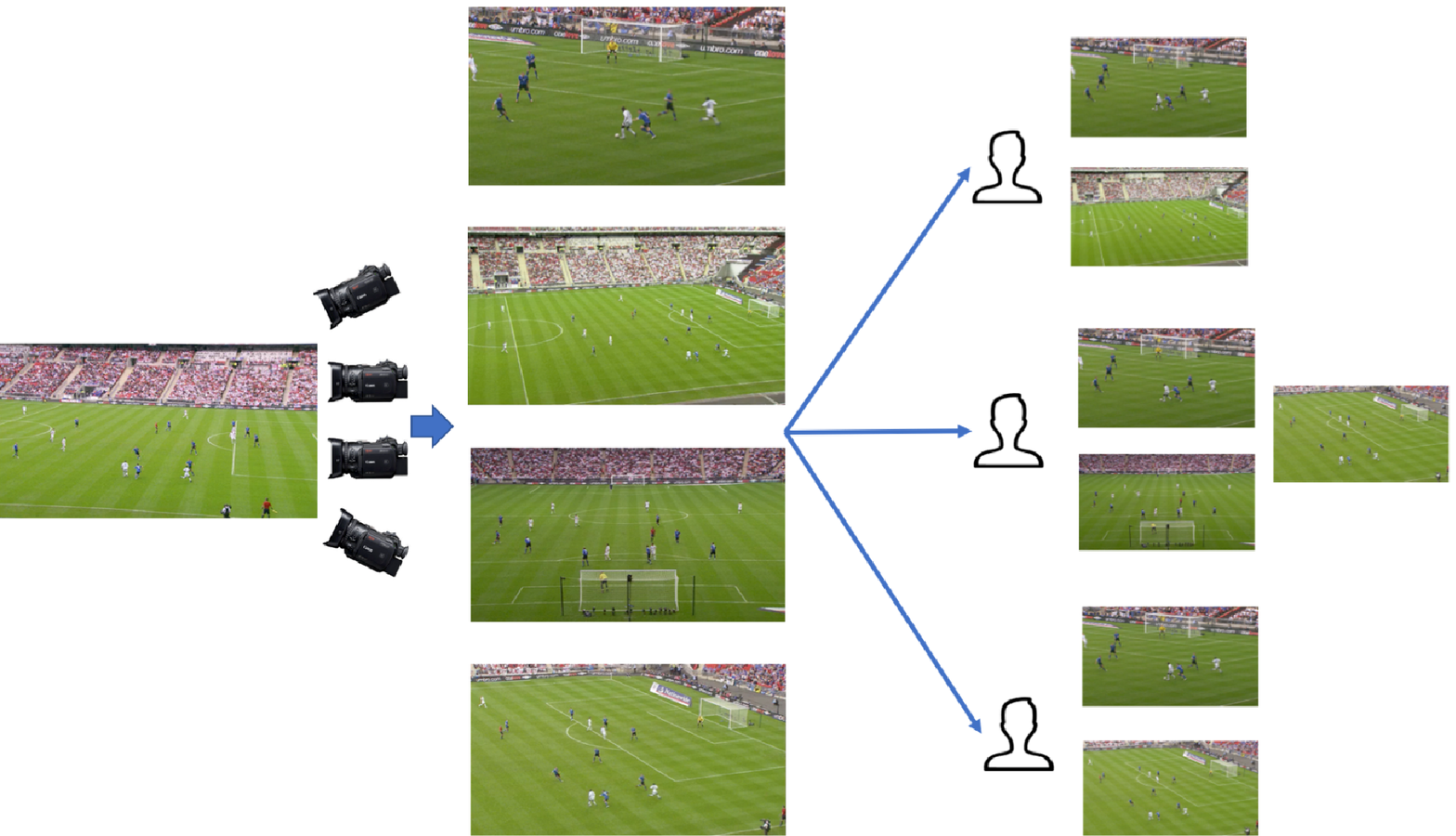}}}
 \end{center}
   \caption{\small{Applications of general multicast.}}
   \label{fig:applications}
\end{figure}

References\cite{TWC21,TCOM20} are pioneer works for supporting wireless streaming of a 360 video\cite{TWC21} and \textcolor{black}{wireless streaming of} a multi-view video\cite{TCOM20}, which are instances of general multicast. Specifically, in \cite{TWC21,TCOM20}, Orthogonal Multiple Access (OMA) is adopted to convert general multicast to per resource block single-group multicast.
\textcolor{black}{While} the OMA-based mechanisms are easy to implement, \textcolor{black}{spatial multiplexing gain is not exploited}. On the other hand, non-orthogonal transmission \textcolor{black}{mechanisms achieve higher} transmission efficiency but \textcolor{black}{are also more challenging due to interference}. Space Division Multiple Access (SDMA) and Non-Orthogonal Multiple Access (NOMA) are \textcolor{black}{two solutions.}
The cost to suppress interference in SDMA can be high when the channels for some users are spatially aligned, \textcolor{black}{while decoding interference in NOMA may not be possible when the interfering message rate is too high.} \textcolor{black}{Thus, SDMA and NOMA may also have unsatisfactory performance.} Rate splitting, originally proposed to effectively support unicast services\cite{TIT1981}, can partially suppress interference and partially decode interference and hence may circumvent the limitations mentioned above.

In \cite{TIT1981,TIT13_yang}, the authors investigate the simplest form of rate splitting for unicast, \textcolor{black}{hereafter} called 1-layer rate splitting, \textcolor{black}{for the two-user interference channel\cite{TIT1981} and two-user multi-antenna broadcast channel\cite{TIT13_yang}, respectively}.
In \cite{TSP16}, \textcolor{black}{the authors investigate the precoder optimization} of 1-layer rate splitting for unicast \textcolor{black}{for Gaussian multiple-input multiple-output channels}.
Later, 1-layer rate splitting for unicast is extended to general rate splitting \textcolor{black}{for unicast}\cite{JSAC21}, 1-layer rate splitting for unicast \textcolor{black}{together} with a multicast message intended for all users \cite{TCOM19}, and multi-group multicast\cite{TVT20}, \textcolor{black}{respectively.} \textcolor{black}{Specifically, \cite{JSAC21,TCOM19,TVT20} focus on the \textcolor{black}{optimizations} of rate splitting with linear beamforming.}  
\textcolor{black}{Optimization-based random linear network coding design for general multicast has been studied in\cite{TON17} for wired networks. Besides general rate splitting for general multicast has been studied in \cite{ISIT17} for discrete memoryless broadcast channels. Here, we are interested in Gaussian fading channels and specifically the linear beamforming design from the optimization perspective.}
Besides, the \textcolor{black}{optimization} of rate splitting with linear beamforming for unicast and its slight generalizations in \cite{TCOM19,TVT20} cannot apply to general multicast. Therefore, for general multicast, the \textcolor{black}{optimization} of general rate splitting with linear beamforming remains an open problem.

This paper \textcolor{black}{intends to shed some light on the above issue}. Specifically, we consider a multi-carrier single-cell wireless network, where a multi-antenna base station (BS) communicates to multiple single-antenna users via general multicast. First, we present general rate splitting for general multicast and characterize the achievable rate regions under linear beamforming at the BS and joint decoding at each user. Then, we optimize the transmission beamforming vectors and rates of sub-message units to maximize the weighted sum rate subject to the achievable rate region constraints and power constraint. \textcolor{black}{Note} that the proposed problem formulation includes those in\cite{JSAC21,TCOM19,TSP06,TVT20} as special cases. This problem is a challenging nonconvex problem. Next, we propose an iterative algorithm to obtain a KKT point using the concave-convex procedure (CCCP). Finally, we numerically demonstrate substantial gains of the proposed solutions over existing schemes and reveal the design insights of general rate splitting for general multicast.

\section{System Model}\label{sec:system}
\textcolor{black}{In this section, we first introduce} general multicast \textcolor{black}{in a single-cell wireless network} and briefly illustrate its connection with \textcolor{black}{unicast, single-group multicast, and multi-group multicast.} \textcolor{black}{Then, we present general rate splitting. Finally, we illustrate the physical layer model and the implementation with linear beamforming and joint decoding.}

\subsection{General Multicast}
We consider a \textcolor{black}{single-cell wireless network consisting of} one BS and $K$ users. Let $\mathcal{K} \triangleq \{1,\ldots,K\}$ denote the set of user indices. \textcolor{black}{The BS has $I$ independent messages.} Let $\mathcal{I}\triangleq \{1,\ldots, I\}$ denote the set of $I$ messages. We consider general multicast. Specifically, each user $k\in\mathcal{K}$ can request arbitrary $I_{k}$ messages in $\mathcal{I}$, denoted by $\mathcal{I}_{k}\subseteq\mathcal{I}$, \textcolor{black}{from the BS. We do not have any assumptions on $\mathcal{I}_{k},~k\in\mathcal{K}$ except that each message in $\mathcal{I}$ is requested by at least one user, i.e., $\cup_{k\in\mathcal{K}}\mathcal{I}_{k} = \mathcal{I}$\cite{TON17}. }

\begin{figure}[t]
\begin{center}
 {\resizebox{9cm}{!}{\includegraphics{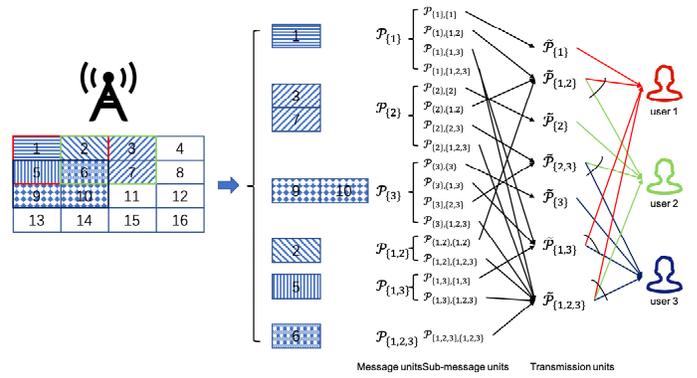}}}
\end{center}
   \caption{\small{\textcolor{black}{Wireless streaming of a tiled 360 video to three users. The 360 video is divided into $4\times 4$ tiles. The users have different FoVs which overlap to certain extent.} $K = 3$, $I = 8$, $\mathcal{I}_{1} = \{1,2,5,6\}$, $\mathcal{I}_{2} = \{2,3,6,7\}$, $\mathcal{I}_{3} = \{5,6,9,10\}$.}}
   \label{system_model}
\end{figure}

\textcolor{black}{To facilitate serving the $K$ users, we partition the message set $\mathcal{I}$ according to the requests from the $K$ users.} For all $\mathcal{S}\subseteq \mathcal{K},\mathcal{S}\neq \emptyset$, let 
\begin{align}
\mathcal{P}_{\mathcal{S}} \triangleq \left(\bigcap_{k\in\mathcal{S}}\mathcal{I}_{k}\right)\bigcap\left(\mathcal{I} - \bigcup_{k\in\mathcal{K}\backslash\mathcal{S}}\mathcal{I}_{k}\right)
\end{align}
denote the set of the messages that is requested by each user in $\mathcal{S}$ and not requested by any user in $\mathcal{K}\backslash\mathcal{S}$ \cite{TWC21}. Define
\begin{align}
\boldsymbol{\mathcal{P}}\triangleq \{\mathcal{P}_{\mathcal{S}}|\mathcal{P}_{\mathcal{S}}\neq \emptyset, \mathcal{S}\subseteq\mathcal{K},\mathcal{S}\neq \emptyset \},\nonumber\\
\boldsymbol{\mathcal{S}} \triangleq \{\mathcal{S}|\mathcal{P}_{\mathcal{S}}\neq \emptyset, \mathcal{S}\subseteq\mathcal{K},\mathcal{S}\neq \emptyset\}.\nonumber
\end{align}
Thus, $\boldsymbol{\mathcal{P}}$ forms a partition of $\mathcal{I}$ and $\boldsymbol{\mathcal{S}}$ specifies the user groups corresponding to the partition. We refer to each element in $\boldsymbol{\mathcal{P}}$ as a message unit.\footnote{\textcolor{black}{$\boldsymbol{\mathcal{P}}$ and $\boldsymbol{\mathcal{S}}$ are assumed to be given in \cite{ISIT17}}.} \textcolor{black}{We can see that different message units in $\boldsymbol{\mathcal{P}}$ are requested by different user groups in $\boldsymbol{\mathcal{S}}$.}
\begin{Exam}[Illustration of $\boldsymbol{\mathcal{P}}$ and $\boldsymbol{\mathcal{S}}$]
As illustrated in Fig. \ref{system_model}, we consider $K = 3$, $I = 8$, $\mathcal{I}_{1} = \{1,2,5,6\}$, $\mathcal{I}_{2} = \{2,3,6,7\}$, $\mathcal{I}_{3} = \{5,6,9,10\}$. Then, we have $\mathcal{P}_{\{1\}} = \{1\}$, $\mathcal{P}_{\{2\}} = \{3,7\}$, $\mathcal{P}_{\{3\}} = \{9,10\}$, $\mathcal{P}_{\{1,2\}} = \{2\}$, $\mathcal{P}_{\{1,3\}} = \{5\}$, $\mathcal{P}_{\{1,2,3\}} = \{6\}$, $\boldsymbol{\mathcal{P}} = \{\mathcal{P}_{\{1\}},\mathcal{P}_{\{2\}},\mathcal{P}_{\{3\}},\mathcal{P}_{\{1,2\}},\mathcal{P}_{\{1,3\}},\mathcal{P}_{\{1,2,3\}}\}$, and $\boldsymbol{\mathcal{S}} = \{\{1\},\{2\},\{3\},\{1,2\},\{1,3\},\{1,2,3\}\}$. There are 6 message units that are requested by 6 groups of users, respectively. For example, message unit $\mathcal{P}_{\{1\}}$ is requested only by user 1, message unit $\mathcal{P}_{\{1,2\}}$ is requested by user 1 and user 2, \textcolor{black}{and} message unit $\mathcal{P}_{\{1,2,3\}}$ is requested by user 1, user 2, and user 3.
\end{Exam}

\begin{Rem}[Connection with Unicast and Multicast]
The considered general multicast includes conventional unicast, single-group multicast, and multi-group multicast as special cases. \textcolor{black}{When $I = K, I_{k} = 1,k\in\mathcal{K}$, and $\mathcal{I}_{k}\not= \mathcal{I}_{k'}, k,k'\in\mathcal{K}, k\not=k'$, general multicast reduces to unicast.} \textcolor{black}{In this case, $\boldsymbol{\mathcal{P}} = \{\{1\},\{2\},\ldots,\{K\}\}$ and $\boldsymbol{\mathcal{S}} = \{\{1\},\{2\},\ldots, \{K\}\}$.} \textcolor{black}{When $I = 1$, implying $I_{k} = 1,k\in\mathcal{K}$, and $\mathcal{I}_{k} = \mathcal{I}_{k'},k,k'\in\mathcal{K},k\not=k'$, general multicast becomes single-group multicast. \textcolor{black}{In this case, $\boldsymbol{\mathcal{P}} = \{\{1\}\}$ and $\boldsymbol{\mathcal{S}} = \{\mathcal{K}\}$}}. When $1< I < K$ and $I_{k} = 1, k\in\mathcal{K}$, general multicast reduces to multi-group ($I$-group) multicast. In this case, $\boldsymbol{\mathcal{P}} = \{\{1\},\{2\},\ldots,\{I\}\}$ and $\boldsymbol{\mathcal{S}} = \left\{\{k\in\mathcal{K}|\mathcal{I}_{k} = \{1\}\},\ldots, \{k\in\mathcal{K}|\mathcal{I}_{k} = \{I\}\}\right\}$. \textcolor{black}{The general multicast considered in this paper, \textcolor{black}{general connection in \cite{TON17}}, and general groupcast considered in \cite{ISIT17} mean the same.}
\end{Rem}

\subsection{General Rate Splitting}\label{sec:rs_model}

We consider rate splitting in the most general form for general multicast to serve the $K$ users\cite{ISIT17}. \textcolor{black}{It allows each user group to decode not only the desired message unit $\mathcal{P}_{\mathcal{S}}$ but also part of the message unit of any other user group, $\mathcal{P}_{\mathcal{S}'}$ for all $\mathcal{S}' \not= \mathcal{S}, \mathcal{S}'\in\boldsymbol{\mathcal{S}}$, to flexibly reduce the interference level.} For all $\mathcal{S}\in\boldsymbol{\mathcal{S}}$, let $\boldsymbol{\mathcal{G}}_{\mathcal{S}} \triangleq \{\mathcal{X}|\mathcal{S}\subseteq\mathcal{X}\subseteq \mathcal{K}\}$.
Namely, $\boldsymbol{\mathcal{G}}_{\mathcal{S}}$ collects all $2^{K - |\mathcal{S}|}$ subsets of $\mathcal{K}$ that contain $\mathcal{S}$. Define $\boldsymbol{\mathcal{G}} \triangleq \bigcup_{\mathcal{S}\in\boldsymbol{\mathcal{S}}} \boldsymbol{\mathcal{G}}_{\mathcal{S}}$. \textcolor{black}{Obviously, $\boldsymbol{\mathcal{S}} \subseteq \boldsymbol{\mathcal{G}}$}.
\textcolor{black}{First, we split each message unit $\mathcal{P}_{\mathcal{S}}$ into \textcolor{black}{$2^{K - |\mathcal{S}|}$} sub-message units, i.e.,
\begin{align}
\mathcal{P}_{\mathcal{S}} = \prod\nolimits_{\mathcal{G}\in\boldsymbol{\mathcal{G}}_{\mathcal{S}}}\mathcal{P}_{\mathcal{S},\mathcal{G}},~\mathcal{S}\in\boldsymbol{\mathcal{S}},
\end{align}
where \textcolor{black}{$\prod$ represents the Cartesian product.} Accordingly,} the rate of the message unit $\mathcal{P}_{\mathcal{S}}$, denoted by $R_{\mathcal{S}}$, is split into the rates of the $2^{K - |\mathcal{S}|}$ sub-message units $\mathcal{P}_{\mathcal{S},\mathcal{G}},\mathcal{G}\in\boldsymbol{\mathcal{G}}_{\mathcal{S}}$,\footnote{When $\mathcal{S} = \mathcal{K}$, $\boldsymbol{\mathcal{G}}_{S} =\{\mathcal{S}\}$ and the message unit $\mathcal{P}_{\mathcal{S}}$ will not be split. For ease of exposition, we let $\mathcal{P}_{\mathcal{S}} = \mathcal{P}_{\mathcal{S},\mathcal{S}}$ and $R_{\mathcal{S}} = R_{\mathcal{S},\mathcal{S}}$ for $\mathcal{S} = \mathcal{K}$.} denoted by $R_{\mathcal{S},\mathcal{G}},\mathcal{G}\in\boldsymbol{\mathcal{G}}_{\mathcal{S}}$ i.e.,
\begin{align}
R_{\mathcal{S}} = \sum\nolimits_{\mathcal{G}\in\boldsymbol{\mathcal{G}}_{\mathcal{S}}}\label{eq:rate_split}
R_{\mathcal{S},\mathcal{G}},~\mathcal{S}\in\boldsymbol{\mathcal{S}}.
\end{align}
\textcolor{black}{Let $\boldsymbol{\mathcal{S}}_{\mathcal{G}} \triangleq \{\mathcal{S}\in\boldsymbol{\mathcal{S}} | \mathcal{S}\subseteq\mathcal{G} \} $}. Then, for all $\mathcal{G}\in\boldsymbol{\mathcal{G}}$, we re-assemble the sub-message units $\mathcal{P}_{\mathcal{S},\mathcal{G}},\mathcal{S}\in\boldsymbol{\mathcal{S}}_{\mathcal{G}}$ to form a transmission unit $\widetilde{\mathcal{P}}_{\mathcal{G}}$ with rate:
\begin{align}
&\widetilde{R}_{\mathcal{G}} = \sum\nolimits_{\mathcal{S}\in\boldsymbol{\mathcal{S}}_{\mathcal{G}}}
R_{\mathcal{S},\mathcal{G}},
~\mathcal{G}\in\boldsymbol{\mathcal{G}}.\label{eq:rate_combine}
\end{align}
\textcolor{black}{That is, we first split $|\boldsymbol{\mathcal{S}}|$ message units, $\mathcal{P}_{\mathcal{S}},\mathcal{S}\in\boldsymbol{\mathcal{S}}$, into $\sum_{\mathcal{S}\in\boldsymbol{\mathcal{S}}}2^{K-|\mathcal{S}|}$ sub-message units, $\mathcal{P}_{\mathcal{S},\mathcal{G}},\mathcal{G}\in\boldsymbol{\mathcal{G}}_{\mathcal{S}},\mathcal{S}\in\boldsymbol{\mathcal{S}}$, and then we re-assemble these sub-message units to form $|\boldsymbol{\mathcal{G}}|$ transmission units, $\widetilde{\mathcal{P}}_{\mathcal{G}},\mathcal{G}\in\boldsymbol{\mathcal{G}}$.}
\begin{Exam}[Illustration of $\boldsymbol{\mathcal{G}}$ and General Rate Splitting]
For Example 1, we have $\mathcal{G}_{\{1\}} = \{\{1\}, \{1,2\}, \{1,3\}, \{1,2,3\}\}$, $\mathcal{G}_{\{2\}} = \{\{2\}, \{1,2\}, \{2,3\}, \{1,2,3\}\}$, $\mathcal{G}_{\{3\}} = \{\{3\}, \{1,3\}, \{2,3\}, \{1,2,3\}\}$, $\mathcal{G}_{\{1,2\}} = \{\{1,2\}, \{1,2,3\}\}$, $\mathcal{G}_{\{1,3\}} = \{\{1,3\}, \{1,2,3\}\}$, $\boldsymbol{\mathcal{G}} = \{\{1\}, \{2\}, \{3\}, \{1,2\}, \{1,3\}, \{2,3\}, \{1,2,3\}\}$. As shown in Fig. \ref{system_model}, 
we first split 6 message units into 17 sub-message units and then re-assemble the 17 sub-message units to form 7 transmission units.
\end{Exam}

\begin{Rem}[Connection with Rate Splitting for Unicast and Multicast]
When general multicast \textcolor{black}{degrades to} unicast, the proposed general rate splitting reduces to the general rate splitting for unicast proposed in our previous work \cite{JSAC21}, which extends \textcolor{black}{the} one-layer rate splitting for unicast \cite{TIT1981}. When general multicast \textcolor{black}{degrades to} single-group multicast, the proposed general rate splitting reduces to the conventional single-group multicast transmission as $\boldsymbol{\mathcal{G}}_{\mathcal{S}} = \boldsymbol{\mathcal{G}} = \{\mathcal{K}\},\mathcal{S}\in\boldsymbol{\mathcal{S}}$. \textcolor{black}{When general multicast degrades to multi-group multicast, the proposed general rate splitting reduces to the one-layer rate splitting for multi-group multicast \cite{TVT20}.}
\end{Rem}

\subsection{Physical Layer Model and Implementation}
The BS is equipped with $M$ antennas, and each user has one antenna. We consider a multi-carrier system. Let $N$ and $\mathcal{N} \triangleq\{1,2, \ldots, N\}$ denote the number of subcarriers and the set of subcarrier indices, respectively. The bandwidth of each subcarrier is $B$ (in Hz). We consider a discrete-time system, i.e., time is divided into fixed-length slots. We adopt the block fading model, i.e., for each user and subcarrier, the channel remains constant within each slot and \textcolor{black}{is independent and identically distributed (i.i.d.) over slots.} \textcolor{black}{We consider slow fading and study an arbitrary slot. 
Let $\mathbf{h}\triangleq (\mathbf{h}_{k,n})_{k\in\mathcal{K},n\in\mathcal{N}}\in\mathbb{C}^{M\times 1}$ denote the system channel state.}
Assume that user $k\in\mathcal{K}$ knows his channel state $\mathbf{h}_{k}\triangleq (\mathbf{h}_{k,n})_{n\in\mathcal{N}}$ and the system channel state $\mathbf{h}$ is known to the BS.

For all $\mathcal{G}\in\boldsymbol{\mathcal{G}}$, transmission unit $\widetilde{\mathcal{P}}_{\mathcal{G}}$ is encoded (channel coding) into codewords that span over the $N$ subcarriers. Let \textcolor{black}{$s_{\mathcal{G},n} \in \mathbb{C}$} denote a symbol for $\widetilde{\mathcal{P}}_{\mathcal{G}}$ that is transmitted on the $n$-th subcarrier. For all $n\in\mathcal{N}$, let $\mathbf{s}_{n} \triangleq (s_{\mathcal{G},n})_{\mathcal{G}\in\boldsymbol{\mathcal{G}}}$, and assume that $\mathbb{E}[\mathbf{s}_{n}\mathbf{s}_{n}^{H}] = \mathbf{I}$. \textcolor{black}{We consider linear beamforming.} \textcolor{black}{For all $n\in\mathcal{N}$,} let $\mathbf{w}_{\mathcal{G},n}\in\mathbb{C}^{M\times 1}$ denote the beamforming vector for \textcolor{black}{transmitting} $\widetilde{\mathcal{P}}_{\mathcal{G}}$ on subcarrier $n$. \textcolor{black}{Using superposition coding,} the transmitted signal on subcarrier $n$, \textcolor{black}{denoted by $\mathbf{x}_{n}\in\mathbb{C}^{M\times 1}$,} is given by:
\begin{align}
\mathbf{x}_{n}=\sum\nolimits_{\mathcal{G}\in\boldsymbol{\mathcal{G}}}
\mathbf{w}_{\mathcal{G},n}s_{\mathcal{G},n},~n\in\mathcal{N}.\label{eq:signal_slow_fading}
\end{align}
The transmission power on subcarrier $n\in\mathcal{N}$ is given by $\sum_{\mathcal{G} \in \boldsymbol{\mathcal{G}}}\| \mathbf{w}_{\mathcal{G},n}\|^2_{2}$, and the total transmission power \textcolor{black}{is given by} $\sum_{n \in \mathcal{N}}\sum_{\mathcal{G} \in \boldsymbol{\mathcal{G}}}\| \mathbf{w}_{\mathcal{G},n} \|^2_{2}$. The total transmission power constraint is given by:
\begin{align}
 \sum_{n \in \mathcal{N}}\sum_{\mathcal{G} \in \boldsymbol{\mathcal{G}}}\| \mathbf{w}_{\mathcal{G},n} \|^2_{2} \leq P. \label{eq:power_perslot}
\end{align}
Here, $P$ denotes the transmission power budget. \textcolor{black}{Define $\boldsymbol{\mathcal{G}}^{(k)} \triangleq \{\mathcal{G} \in \boldsymbol{\mathcal{G}} | k\in\mathcal{G} \}, k\in\mathcal{K}$.}
Then, the received signal at user $k\in\mathcal{K}$ on subcarrier $n\in\mathcal{N}$, \textcolor{black}{denoted by $y_{k,n}\in\mathbb{C}$}, is given by:
\begin{align}
y_{k,n}& = \mathbf{h}^{H}_{k,n}\mathbf{x}_{n} + z_{k,n} =  \mathbf{h}_{k,n}^H \sum\nolimits_{\mathcal{G}\in\boldsymbol{\mathcal{G}}^{(k)}}
\mathbf{w}_{\mathcal{G},n}s_{\mathcal{G},n}\nonumber\\
&+\mathbf{h}_{k,n}^H\sum\nolimits_{\mathcal{G}'\in\boldsymbol{\mathcal{G}}\backslash\boldsymbol{\mathcal{G}}^{(k)}} \mathbf{w}_{\mathcal{G}',n}s_{\mathcal{G}',n} +z_{k,n},\nonumber\\
&~~~~~~~~~~~~~~~~~~~~~~~~~~~~~~~~~~~~~~k \in \mathcal{K},~n\in\mathcal{N},\label{eq:receive_signal}
\end{align}
where \textcolor{black}{the last equality is due to \eqref{eq:signal_slow_fading}, and }$z_{k,n}\sim C\mathcal{N}(0,\sigma^2)$ is the additive white gaussian noise (AWGN). \textcolor{black}{In \eqref{eq:receive_signal}, the first term represents the desired signal, and the second represents the interference.} \textcolor{black}{It is noteworthy that} the main idea of rate splitting is to make the \textcolor{black}{undesired} messages partially decodable in order to reduce interference \cite{JSAC21}.
To exploit the full potential of the general rate splitting for general multicast, we consider joint decoding \textcolor{black}{at each user}.\footnote{We can easily extend it to successive decoding as in \cite{JSAC21}.} That is, each user $k\in\mathcal{K}$ jointly decodes the desired transmission units $\widetilde{\mathcal{P}}_{\mathcal{G}},\mathcal{G}\in\boldsymbol{\mathcal{G}}^{(k)}$. Thus, the achievable rate region of the transmission units is described by the following constraints:
\begin{align}
\sum\limits_{\mathcal{G}\in\boldsymbol{\mathcal{X}}}&\widetilde{R}_{\mathcal{G}} \nonumber\\
&\leq B\sum\limits_{n\in\mathcal{N}}\log_{2}\left(1+\frac{\sum\nolimits_{\mathcal{G}\in\boldsymbol{\mathcal{X}}}|\mathbf{h}^{H}_{k,n}\mathbf{w}_{\mathcal{G},n}|^{2}}{\sigma^2 +\sum\nolimits_{\mathcal{G}'\in\boldsymbol{\mathcal{G}}\backslash\boldsymbol{\mathcal{G}}^{(k)}}|\mathbf{h}^{H}_{k,n}\mathbf{w}_{\mathcal{G}^{'},n}|^{2} }\right),\nonumber\\
&~~~~~~~~~~~~~~~~~~~~~~~~~~~~~~~~~~~~~\boldsymbol{\mathcal{X}}\subseteq\boldsymbol{\mathcal{G}}^{(k)},k \in \mathcal{K},\label{eq:rateconstraints_perslot}
\end{align}
\textcolor{black}{where $\widetilde{R}_{\mathcal{G}}$ is given by \eqref{eq:rate_combine}.} 

\section{Optimization Problem Formulation}\label{sec:slowfading}
In this section, we would like to optimize the transmission beamforming vectors $\mathbf{w} \triangleq (\mathbf{w}_{\mathcal{G},n})_{\mathcal{G}\in\boldsymbol{\mathcal{G}},n\in\mathcal{N}}$ and rates of the sub-message units $\mathbf{R} \triangleq (R_{\mathcal{S},\mathcal{G}})_{\mathcal{S}\in\boldsymbol{\mathcal{S}}, \mathcal{G}\in\boldsymbol{\mathcal{G}}}$ to maximize the \textcolor{black}{weighted sum rate},\footnote{\textcolor{black}{The proposed problem formulation and solution method can be readily extended to maximize the sum rate and worst-case rate as in \cite{JSAC21}.}} $\sum\nolimits_{\mathcal{S}\in\boldsymbol{\mathcal{S}}} \alpha_{\mathcal{S}} R_{\mathcal{S}}$, \textcolor{black}{where the coefficient $\alpha_{\mathcal{S}} \geq 0$ denotes the weight for message unit $\mathcal{P}_{\mathcal{S}}$, subject to the total transmission power constraint in \eqref{eq:power_perslot} and the achievable rate constraints in \eqref{eq:rateconstraints_perslot}.}
Therefore, we formulate the following optimization problem.

\begin{Prob}[Weighted Sum Rate Maximization]\label{prob:UMwT_LMsg_1slot_decouple}
\begin{align}
\max_{\mathbf{w},\mathbf{R}\succeq 0}&\quad 
\sum\limits_{\mathcal{S}\in\boldsymbol{\mathcal{S}}} \alpha_{\mathcal{S}} R_{\mathcal{S}}\nonumber\\
\mathrm{s.t.}\quad &~\eqref{eq:power_perslot},~\eqref{eq:rateconstraints_perslot}. \nonumber
\end{align}
\end{Prob}

\begin{Rem}[Connection with Rate Splitting for Unicast and Multicast]
When general multicast degrades to unicast, Problem \ref{prob:UMwT_LMsg_1slot_decouple} reduces to the weighted sum rate maximization problem \textcolor{black}{for general rate splitting for unicast} in \cite{JSAC21}. When general multicast degrades to single-group multicast, Problem~\ref{prob:UMwT_LMsg_1slot_decouple} reduces to the rate maximization problem for single-group multicast in \cite{TSP06}. \textcolor{black}{Finally, when general multicast degrades to multi-group multicast, Problem~\ref{prob:UMwT_LMsg_1slot_decouple} \textcolor{black}{can be viewed as a generalization of} the weighted sum rate maximization for multi-group multicast in \cite{TVT20}. }
\end{Rem}

Note that the objective function is linear, the constraint in \eqref{eq:power_perslot} is convex, and the constraints in \eqref{eq:rateconstraints_perslot} are nonconvex. Thus, Problem~\ref{prob:UMwT_LMsg_1slot_decouple} is nonconvex.\footnote{\textcolor{black}{There are generally no effective methods for solving a nonconvex problem optimally. The goal of solving a nonconvex problem is usually to design an iterative algorithm to obtain a stationary point or a KKT point (which satisfies necessary conditions for optimality if strong
duality holds).}}

\section{Solution}
In this section, we propose an iterative algorithm to obtain a KKT point of Problem~\ref{prob:UMwT_LMsg_1slot_decouple} using CCCP. 
First, we transform Problem~\ref{prob:UMwT_LMsg_1slot_decouple} into the following equivalent problem by introducing \textcolor{black}{auxiliary variables $\mathbf{e} \triangleq \left(e_{k,n,\boldsymbol{\mathcal{X}}}\right)
_{\boldsymbol{\mathcal{X}}\subseteq\boldsymbol{\mathcal{G}}^{(k)},k\in\mathcal{K},n\in\mathcal{N}}$ and
$\mathbf{u} \triangleq \left(u_{k,n,\boldsymbol{\mathcal{X}}}\right)_{\boldsymbol{\mathcal{X}}\subseteq\boldsymbol{\mathcal{G}}^{(k)},k\in\mathcal{K},n\in\mathcal{N}}$ and extra constraints:}
\begin{align}
& \sum\limits_{\mathcal{G}'\in\boldsymbol{\mathcal{G}}\backslash\boldsymbol{\mathcal{G}}^{(k)}}|\mathbf{h}^{H}_{k,n}\mathbf{w}_{\mathcal{G}^{'},n}|^{2}
+\sigma^2 \nonumber\\
&- \frac{\sum\limits_{\mathcal{G}\in\boldsymbol{\mathcal{X}}}|
\mathbf{h}^{H}_{k,n}\mathbf{w}_{\mathcal{G},n}|^{2}+
\sum\limits_{\mathcal{G}'\in\boldsymbol{\mathcal{G}}\backslash\boldsymbol{\mathcal{G}}^{(k)}}|\mathbf{h}^{H}_{k,n}\mathbf{w}_{\mathcal{G}^{'},n}|^{2}
+\sigma^2}{u_{k,n,\boldsymbol{\mathcal{X}}}(\mathbf{h})} \leq 0,\nonumber\\
&~~~~~~~~~~~~~~~~~~~~~~~~~~~~~~~~~\boldsymbol{\mathcal{X}}\subseteq\boldsymbol{\mathcal{G}}^{(k)},k\in\mathcal{K},n\in\mathcal{N},\label{eq:equivalentDCfunction_perslot}
\end{align}
\begin{align}
&\sum_{\mathcal{G}\in\boldsymbol{\mathcal{X}}}\sum_{\mathcal{S}\in\boldsymbol{\mathcal{S}}_{\mathcal{G}}}
R_{\mathcal{S},\mathcal{G}}=\sum_{n\in\mathcal{N}} e_{k,n,\boldsymbol{\mathcal{X}}},~\boldsymbol{\mathcal{X}}\subseteq\boldsymbol{\mathcal{G}}^{(k)},k\in\mathcal{K},\label{eq:DC_R<e_perslot}\\ 
& 2^{\frac{e_{k,n,\boldsymbol{\mathcal{X}}}}{B}} \leq  u_{k,n,\boldsymbol{\mathcal{X}}},~\boldsymbol{\mathcal{X}}\subseteq\boldsymbol{\mathcal{G}}^{(k)},k\in\mathcal{K},n\in\mathcal{N}. \label{eq:DC_e<u_perslot}
\end{align}

\begin{Prob}[Equivalent Problem of Problem \ref{prob:UMwT_LMsg_1slot_decouple}]\label{prob:UMwT_LMsg_1slot_decouple_DC}
\begin{align}
\max_{\mathbf{w},\mathbf{R}\succeq 0,\mathbf{e},\mathbf{u}}\quad &\sum\limits_{\mathcal{S}\in\boldsymbol{\mathcal{S}}} \alpha_{\mathcal{S}} \sum_{\mathcal{G}\in\boldsymbol{\mathcal{G}}_{\mathcal{S}}}R_{\mathcal{S},\mathcal{G}}  \nonumber\\
\mathrm{s.t.}\quad &\eqref{eq:power_perslot},~\eqref{eq:equivalentDCfunction_perslot},~\eqref{eq:DC_R<e_perslot},~\eqref{eq:DC_e<u_perslot}.\nonumber
\end{align}
Let $(\mathbf{w}^{\star}, \mathbf{R}^{\star}, \mathbf{e}^{\star}, \mathbf{u}^{\star})$ denote an optimal solution of Problem~\ref{prob:UMwT_LMsg_1slot_decouple_DC}.
\end{Prob}

\begin{Thm}[Equivalence Between Problem~\ref{prob:UMwT_LMsg_1slot_decouple} and Problem~\ref{prob:UMwT_LMsg_1slot_decouple_DC}]\label{lemma_DC_transform}
$(\mathbf{w}^{\star}, \mathbf{R}^{\star}, \mathbf{e}^{\star}, \mathbf{u}^{\star})$ satisfies
\textcolor{black}{$2^{\frac{e^{\star}_{k,n,\boldsymbol{\mathcal{X}}}}{B}} = u^{\star}_{k,n,\boldsymbol{\mathcal{X}}},~\boldsymbol{\mathcal{X}}\subseteq\boldsymbol{\mathcal{G}}^{(k)},k\in\mathcal{K},n\in\mathcal{N}.$}
Furthermore, Problem~\ref{prob:UMwT_LMsg_1slot_decouple} and Problem~\ref{prob:UMwT_LMsg_1slot_decouple_DC} are equivalent.
\end{Thm}
\begin{Proof}
First, by introducing auxiliary variables $\mathbf{e}$ and $\mathbf{u}$ and extra constraints:
\begin{align}
2^{\frac{e_{k,n,\boldsymbol{\mathcal{X}}}}{B}} =  u_{k,n,\boldsymbol{\mathcal{X}}},~\boldsymbol{\mathcal{X}}\subseteq\boldsymbol{\mathcal{G}}^{(k)},k\in\mathcal{K},n\in\mathcal{N}, \label{eq:DC_e<u_perslot_proof}
\end{align}
we can equivalently transform Problem~\ref{prob:UMwT_LMsg_1slot_decouple} into the following problem:
\begin{align}
\max_{\mathbf{w},\mathbf{R}\succeq 0,\mathbf{e},\mathbf{u}}\quad &\sum\limits_{\mathcal{S}\in\boldsymbol{\mathcal{S}}} \alpha_{\mathcal{S}} \sum_{\mathcal{G}\in\boldsymbol{\mathcal{G}}_{\mathcal{S}}}R_{\mathcal{S},\mathcal{G}}  \nonumber\\
\mathrm{s.t.}\quad &\eqref{eq:power_perslot},~\eqref{eq:equivalentDCfunction_perslot},~\eqref{eq:DC_R<e_perslot},~\eqref{eq:DC_e<u_perslot_proof}.\nonumber
\end{align}
Let $(\mathbf{w}^{\ddag},\mathbf{R}^{\ddag},\mathbf{e}^{\ddag}(\mathbf{h}),\mathbf{u}^{\ddag})$ denote an optimal solution.
It is obvious that $(\mathbf{w}^{\ddag},\mathbf{R}^{\ddag})$ is an optimal solution of Problem \ref{prob:UMwT_LMsg_1slot_decouple}. Next, we transform the above problem to Problem~\ref{prob:UMwT_LMsg_1slot_decouple_DC} by relaxing the constrains in \eqref{eq:DC_e<u_perslot_proof} to the constraints in \eqref{eq:DC_e<u_perslot}. By contradiction and the monotonicity of the objective function with respect to (w.r.t.) $\mathbf{R}$ in Problem~\ref{prob:UMwT_LMsg_1slot_decouple_DC}, we can show that the constraints in \eqref{eq:DC_e<u_perslot} are active at the optimal solution. Thus, $(\mathbf{w}^{\ddag},\mathbf{R}^{\ddag},\mathbf{e}^{\ddag},\mathbf{u}^{\ddag})$ is an optimal solution of Problem~\ref{prob:UMwT_LMsg_1slot_decouple_DC}. Therefore, we can show Theorem~\ref{lemma_DC_transform}.$\hfill\blacksquare$
\end{Proof}
\begin{figure*}[ht]
\small{
\begin{align}
&L_{k,n,\boldsymbol{\mathcal{X}}}(\mathbf{w}_{n}, u_{k,n,\boldsymbol{\mathcal{X}}};\mathbf{w}^{(i-1)}_{n}, u^{(i-1)}_{k,n,\boldsymbol{\mathcal{X}}}) \triangleq \sum\limits_{\mathcal{G}'\in\boldsymbol{\mathcal{G}}\backslash\boldsymbol{\mathcal{G}}^{(k)}}|\mathbf{h}^{H}_{k,n}\mathbf{w}_{\mathcal{G}^{'},n}|^{2}
+\sigma^2 + \frac{\left(\sum\limits_{\mathcal{G}\in\boldsymbol{\mathcal{X}}}
|\mathbf{h}^{H}_{k,n}\mathbf{w}^{(i-1)}_{\mathcal{G},n}|^{2}+
\sum\limits_{\mathcal{G}'\in\boldsymbol{\mathcal{G}}\backslash\boldsymbol{\mathcal{G}}^{(k)}}|\mathbf{h}^{H}_{k,n}\mathbf{w}^{(i-1)}_{\mathcal{G}^{'},n}|^{2}
+\sigma^2
\right)u_{k,n,\boldsymbol{\mathcal{X}}}}{\left(u^{(i-1)}_{k,n,\boldsymbol{\mathcal{X}}}\right)^2} \nonumber\\
&- \frac{2\Re\left\{\sum\limits_{\mathcal{G}\in\boldsymbol{\mathcal{X}}}
\mathbf{w}_{\mathcal{G},n}^{(i-1)H}\mathbf{h}_{k,n}\mathbf{h}^{H}_{k,n}
\mathbf{w}_{\mathcal{G},n}+\sum\limits_{\mathcal{G}'\in\boldsymbol{\mathcal{G}}\backslash\boldsymbol{\mathcal{G}}^{(k)}}
\mathbf{w}_{\mathcal{G}^{'},n}^{(i-1)H}\mathbf{h}_{k,n}\mathbf{h}^{H}_{k,n}
\mathbf{w}_{\mathcal{G}^{'},n}\right\}
+2\sigma^2}{u^{(i-1)}_{k,n,\boldsymbol{\mathcal{X}}}},~\boldsymbol{\mathcal{X}}\subseteq\boldsymbol{\mathcal{G}}^{(k)},k\in\mathcal{K},n\in\mathcal{N}.\label{eq:iter_define}
\end{align}
}
\hrulefill
\end{figure*}

Based on Theorem~\ref{lemma_DC_transform}, solving Problem~\ref{prob:UMwT_LMsg_1slot_decouple} is equivalent to solving Problem~\ref{prob:UMwT_LMsg_1slot_decouple_DC}. \textcolor{black}{Problem~\ref{prob:UMwT_LMsg_1slot_decouple_DC} is a difference of convex functions (DC) programming (one type of nonconvex problems) and a KKT point can be obtained by CCCP\textcolor{black}{\cite{JSAC21}}.\footnote{CCCP can exploit the partial convexity and usually converges faster to a KKT point than conventional gradient methods.} The main idea of CCCP is to solve a sequence of successively refined approximate convex problems, each of which is obtained by linearizing the concave part and preserving the remaining convex part in the DC problem. Specifically, at the $i$-th iteration, the approximate convex problem of Problem~\ref{prob:UMwT_LMsg_1slot_decouple_DC} is given as follows. Let $(\mathbf{w}^{(i)},\mathbf{R}^{(i)}, \mathbf{e}^{(i)},\mathbf{u}^{(i)})$ denote an optimal solution of the following problem.}


\begin{Prob}[Approximation of Problem~\ref{prob:UMwT_LMsg_1slot_decouple_DC} at Iteration $i$]\label{prob:UMwT_LMsg_1slot_decouple_DC_approxi}
\begin{align}
&\max_{\mathbf{w},\mathbf{R}\succeq 0,\mathbf{e},\mathbf{u}}\quad \sum\limits_{\mathcal{S}\in\boldsymbol{\mathcal{S}}} \alpha_{\mathcal{S}} \sum_{\mathcal{G}\in\boldsymbol{\mathcal{G}}_{\mathcal{S}}}R_{\mathcal{S},\mathcal{G}}  \nonumber\\
&\mathrm{s.t.}\quad~\eqref{eq:power_perslot},~\eqref{eq:DC_R<e_perslot},~\eqref{eq:DC_e<u_perslot},\nonumber\\
&L_{k,n,\boldsymbol{\mathcal{X}}}(\mathbf{w}_{n}, u_{k,n,\boldsymbol{\mathcal{X}}};\mathbf{w}^{(i-1)}_{n}, u^{(i-1)}_{k,n,\boldsymbol{\mathcal{X}}}) \leq 0,\nonumber\\
&~~~~~~~~~~~~~~~~~~~~~~~~~~~~~\boldsymbol{\mathcal{X}}\subseteq\boldsymbol{\mathcal{G}}^{(k)},k\in\mathcal{K},n\in\mathcal{N},\label{eq:DC_approximation_1slot}
\end{align}
where $\mathbf{w}^{(i-1)}_{n} \triangleq (\mathbf{w}^{(i-1)}_{\mathcal{G},n})_{\mathcal{G}\in\boldsymbol{\mathcal{X}} \cup (\boldsymbol{\mathcal{G}}\backslash\boldsymbol{\mathcal{G}}^{(k)})}$, $\mathbf{w}_{n} \triangleq (\mathbf{w}_{\mathcal{G},n})_{\mathcal{G}\in\boldsymbol{\mathcal{X}} \cup (\boldsymbol{\mathcal{G}}\backslash\boldsymbol{\mathcal{G}}^{(k)})}$, and $L_{k,n,\boldsymbol{\mathcal{X}}}(\mathbf{w}_{n}, u_{k,n,\boldsymbol{\mathcal{X}}};\mathbf{w}^{(i-1)}_{n}, u^{(i-1)}_{k,n,\boldsymbol{\mathcal{X}}})$ is given by \eqref{eq:iter_define}, as shown at the top of this page.
\end{Prob}

Problem \ref{prob:UMwT_LMsg_1slot_decouple_DC_approxi} is convex and can be solved efficiently using standard convex optimization methods. Problem~\ref{prob:UMwT_LMsg_1slot_decouple_DC_approxi} has $MN|\boldsymbol{\mathcal{G}}| + \sum_{\mathcal{S}\in\boldsymbol{\mathcal{S}}}2^{K-|\mathcal{S}|} + 2N\sum_{k\in\mathcal{K}} (2^{|\boldsymbol{\mathcal{G}}^{(k)}|} -1 )$ variables and $1 + (2N + 1)\sum_{k\in\mathcal{K}} (2^{|\boldsymbol{\mathcal{G}}^{(k)}|} -1 )$ constraints. Thus, when an interior point method is applied, the computational complexity for solving Problem \ref{prob:UMwT_LMsg_1slot_decouple_DC_approxi} is \textcolor{black}{$\mathcal{O}(K^{3.5}2^{1.75 \times 2^{K}})$ as $K \rightarrow \infty$.} The details of CCCP for obtaining a KKT point of Problem~\ref{prob:UMwT_LMsg_1slot_decouple_DC} are summarized in Algorithm~\ref{alg:CCCP_perslot}.\footnote{In practice, we can run Algorithm 1 multiple times with different feasible initial points to obtain multiple KKT points and choose the KKT point with the best objective value.} As the number of iterations of Algorithm~\ref{alg:CCCP_perslot} does not scale with the problem size\cite{MOR21}, the computational complexity for Algorithm~\ref{alg:CCCP_perslot} is the same as that for solving Problem \ref{prob:UMwT_LMsg_1slot_decouple_DC_approxi}, i.e., \textcolor{black}{$\mathcal{O}(K^{3.5}2^{1.75 \times 2^{K}})$ as $K \rightarrow \infty$.}

\begin{Thm}[Convergence of Algorithm~\ref{alg:CCCP_perslot}]\label{claim_DC_perslot}
As $i \rightarrow \infty$, $\left(\mathbf{w}^{(i)},\mathbf{R}^{(i)}, \mathbf{e}^{(i)}, \mathbf{u}^{(i)}\right)$ obtained by Algorithm~\ref{alg:CCCP_perslot} converges to a KKT point of Problem \ref{prob:UMwT_LMsg_1slot_decouple_DC} \cite{TSP17}.
\end{Thm}

\begin{Proof}
The constraints in \eqref{eq:power_perslot}, \eqref{eq:DC_R<e_perslot}, \eqref{eq:DC_e<u_perslot} are convex, and the constraint function in \eqref{eq:equivalentDCfunction_perslot} can be regarded as a difference between two convex functions, i.e., $\sum\nolimits_{\mathcal{G}'\in\boldsymbol{\mathcal{G}}\backslash\boldsymbol{\mathcal{G}}^{(k)}}|\mathbf{h}^{H}_{k,n}\mathbf{w}_{\mathcal{G}^{'},n}|^{2} +\sigma^2$ and $\frac{\sum\nolimits_{\mathcal{G}\in\boldsymbol{\mathcal{X}}}|
\mathbf{h}^{H}_{k,n}\mathbf{w}_{\mathcal{G},n}|^{2}+
\sum\nolimits_{\mathcal{G}'\in\boldsymbol{\mathcal{G}}\backslash\boldsymbol{\mathcal{G}}^{(k)}}|\mathbf{h}^{H}_{k,n}\mathbf{w}_{\mathcal{G}^{'},n}|^{2}
+\sigma^2}{u_{k,n,\boldsymbol{\mathcal{X}}}}$. Therefore, Problem~\ref{prob:UMwT_LMsg_1slot_decouple_DC} is a DC programming. 
\textcolor{black}{Linearizing $\frac{\sum\nolimits_{\mathcal{G}\in\boldsymbol{\mathcal{X}}}|\mathbf{h}^{H}_{k,n}\mathbf{w}_{\mathcal{G},n}|^{2}+\sum\nolimits_{\mathcal{G}'\in\boldsymbol{\mathcal{G}}\backslash\boldsymbol{\mathcal{G}}^{(k)}}|\mathbf{h}^{H}_{k,n}\mathbf{w}_{\mathcal{G}^{'},n}|^{2}+\sigma^2}{u_{k,n,\boldsymbol{\mathcal{X}}}}$ at $(\mathbf{w}^{(i-1)}_{n}, u^{(i-1)}_{k,n,\boldsymbol{\mathcal{X}}})$ and preserving $\sum\nolimits_{\mathcal{G}'\in\boldsymbol{\mathcal{G}}\backslash\boldsymbol{\mathcal{G}}^{(k)}}|\mathbf{h}^{H}_{k,n}\mathbf{w}_{\mathcal{G}^{'},n}|^{2} +\sigma^2$ give $L_{k,n,\boldsymbol{\mathcal{X}}}(\mathbf{w}_{n}, u_{k,n,\boldsymbol{\mathcal{X}}};\mathbf{w}^{(i-1)}_{n}, u^{(i-1)}_{k,n,\boldsymbol{\mathcal{X}}})$ in \eqref{eq:iter_define}. Thus, Algorithm~1 implements CCCP.}
It has been validated in \cite{TSP17} that solving DC programming through CCCP always returns a KKT point. Therefore, we can show Theorem~\ref{claim_DC_perslot}.$\hfill\blacksquare$
\end{Proof}

\textcolor{black}{This paper focuses mainly on exploiting the full potential of general rate splitting for general multicast. The computational complexity of Algorithm~1 can be formidable with a large $K$. We can use successive decoding to reduce the computational complexity to $\mathcal{O}(K^{1.5}|\boldsymbol{\mathcal{S}}|^{1.5}2^{2K})$ as in \cite{JSAC21}.\footnote{\textcolor{black}{Note that $|\boldsymbol{\mathcal{S}}|$ may not scale with $K$ and how $|\boldsymbol{\mathcal{S}}|$ scales with $K$ relies on the user requests. In the case of $|\boldsymbol{\mathcal{S}}| = \mathcal{O}(1)$, the reduced computational complexity is $\mathcal{O}(K^{1.5}2^{2K})$, as $K \rightarrow \infty$.}} We can also apply rate splitting with a reduced number of layers together with successive decoding to further reduce the computational complexity to $\mathcal{O}(K^{1.5}|\boldsymbol{\mathcal{S}}|^{1.5}(|\boldsymbol{\mathcal{G}}_{\text{lb}}|^{2} + |\boldsymbol{\mathcal{G}}_{\text{lb}}||\boldsymbol{\mathcal{S}}|K + |\boldsymbol{\mathcal{S}}|^{2}K^{2}))$ as in\cite{JSAC21}, for some $\boldsymbol{\mathcal{G}}_{\text{lb}}$ satisfying $\boldsymbol{\mathcal{S}}\subseteq \boldsymbol{\mathcal{G}}_{\text{lb}} \subseteq \boldsymbol{\mathcal{G}}.$ Note that $|\boldsymbol{\mathcal{G}}_{\text{lb}}|$ represents the reduced number of layers and satisfies $|\boldsymbol{\mathcal{S}}| \leq |\boldsymbol{\mathcal{G}}_{\text{lb}}| \leq 2^{K} - 1$.\footnote{In the case of $|\boldsymbol{\mathcal{S}}| = \mathcal{O}(1)$ and $|\boldsymbol{\mathcal{G}}_{\text{lb}}| = \mathcal{O}(1)$, the reduced computational complexity is $\mathcal{O}(N^{3.5}K^{3.5})$ as $K \rightarrow \infty$.} Low-complexity optimization methods are beyond the scope of this paper.}

\begin{algorithm}[t] \caption{Obtaining a KKT Point of Problem~\ref{prob:UMwT_LMsg_1slot_decouple_DC}}
\small{\begin{algorithmic}[1]
\STATE Initialization: Choose any feasible point of Problem~\ref{prob:UMwT_LMsg_1slot_decouple_DC} $(\mathbf{w}^{(0)},\mathbf{R}^{(0)}, \mathbf{e}^{(0)},\mathbf{u}^{(0)})$ and set $i=0$.
\STATE \textbf{repeat}
\STATE Obtain an optimal solution $(\mathbf{w}^{(i)},\mathbf{R}^{(i)}, \mathbf{e}^{(i)},\mathbf{u}^{(i)})$ of Problem~\ref{prob:UMwT_LMsg_1slot_decouple_DC_approxi} with an interior point method. 
\STATE Set $i=i+1$.
\STATE \textbf{until} the convergence criterion $\|(\mathbf{w}^{(i)},\mathbf{R}^{(i)}, \mathbf{e}^{(i)},\mathbf{u}^{(i)}) - (\mathbf{w}^{(i-1)},\mathbf{R}^{(i-1)}, \mathbf{e}^{(i-1)},\mathbf{u}^{(i-1)})\|_{2} \leq \epsilon$ is met.
\end{algorithmic}}\normalsize\label{alg:CCCP_perslot}
\end{algorithm}

\section{Numerical Results}\label{sec:simulation}

\begin{figure}[t]
\begin{center}
 {\resizebox{4cm}{!}{\includegraphics{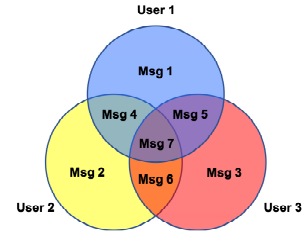}}}
\end{center}
   \caption{\small{General multicast setup in Section~\ref{sec:simulation}.}}
   \label{fig:msg}
\end{figure}

In this section, we numerically evaluate \textcolor{black}{the proposed solution obtained by} Algorithm~\ref{alg:CCCP_perslot}, namely Prop-RS. We consider three baseline schemes, namely 1L-RS, NoRS, and OFDMA. 1L-RS and NoRS extend 1-layer rate splitting\cite{TSP16} and SDMA\cite{TCOM07}, \textcolor{black}{both} for unicast, to general multicast. More specifically, 1L-RS and NoRS implement Algorithm~\ref{alg:CCCP_perslot} to obtain KKT points of Problem~\ref{prob:UMwT_LMsg_1slot_decouple} with $\mathcal{G}_{\mathcal{S}} = \{\mathcal{S},\mathcal{K}\}, \mathcal{S}\in\boldsymbol{\mathcal{S}}$ and with $\mathcal{G}_{\mathcal{S}} = \{\mathcal{S}\}, \mathcal{S}\in\boldsymbol{\mathcal{S}}$, respectively. 
OFDMA considers the maximum ratio transmission (MRT) on each subcarrier and optimizes the subcarrier and power allocation\cite{TWC21}. 

In the simulation, we set $K = 3$, $I = 7$, $I_{1} = \{1,4,5,7\}$, $I_{2} = \{2,4,6,7\}$, and $I_{3} = \{3,5,6,7\}$, as illustrated in Fig. \ref{fig:msg}. As a result, we have $\mathcal{P}_{\{1\}} = \{1\}$, $\mathcal{P}_{\{2\}} = \{2\}$, $\mathcal{P}_{\{3\}} = \{3\}$, $\mathcal{P}_{\{1,2\}} = \{4\}$, $\mathcal{P}_{\{1,3\}} = \{5\}$, $\mathcal{P}_{\{2,3\}} = \{6\}$, and $\mathcal{P}_{\{1,2,3\}} = \{7\}$. Additionally, we set $\alpha_{\mathcal{S}} = 1/7,\mathcal{S}\in\boldsymbol{\mathcal{S}}$, $B$ = 30 kHz, $N$ = 72, and $\sigma^{2} = 10^{-9}$ W. We consider spatially
correlated channel with the correlation following the one-ring scattering model as in \cite{JSAC21}. When applying the one-ring scattering model, let $G$ denote the number of user groups. We set the same angular spreads for the $G$ groups and the same azimuth angle for the users in each group as in\cite{JSAC21}. Note that $G$ is related to the channel correlation among users. Specifically, the correlation increases as $G$ decreases. \textcolor{black}{When $G = 1$, all users belong to one group and have the same channel covariance matrix. When $G = 3$, all users are in different groups and have different channel covariance matrices.} We generate 100 realizations of random system channel state, solve the weighted sum rate maximization problem for each realization, and evaluate \textcolor{black}{the average of} the weighted sum rate of each scheme over the 100 random realizations.

\begin{figure}
  \centering
  \begin{minipage}{.45\columnwidth}
    \centering
    \includegraphics[width=\textwidth]{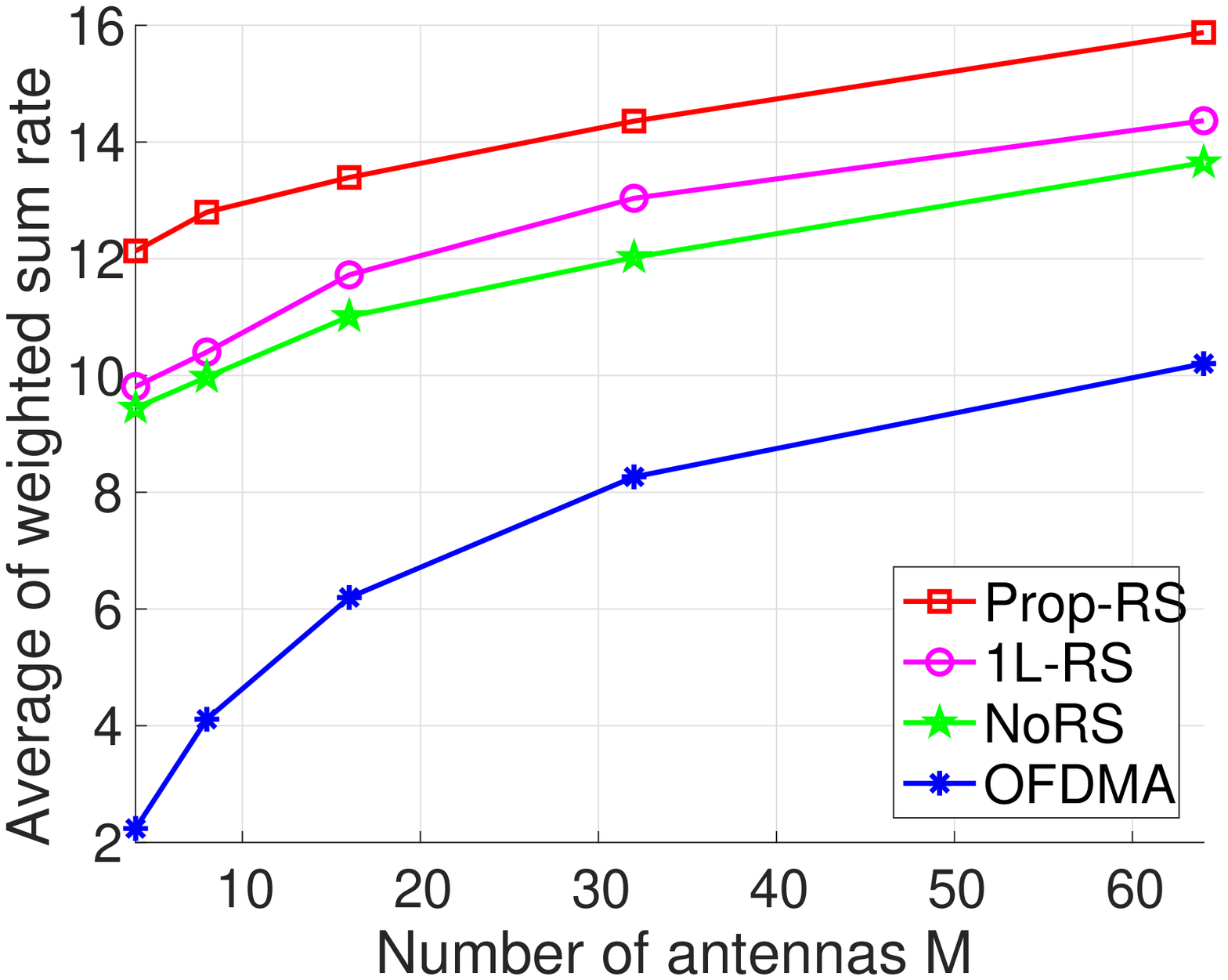}
    \caption{Weighted sum rate versus $M$.}
    \label{fig:label1}
  \end{minipage}%
  \begin{minipage}{.45\columnwidth}
    \centering
    \includegraphics[width=\textwidth]{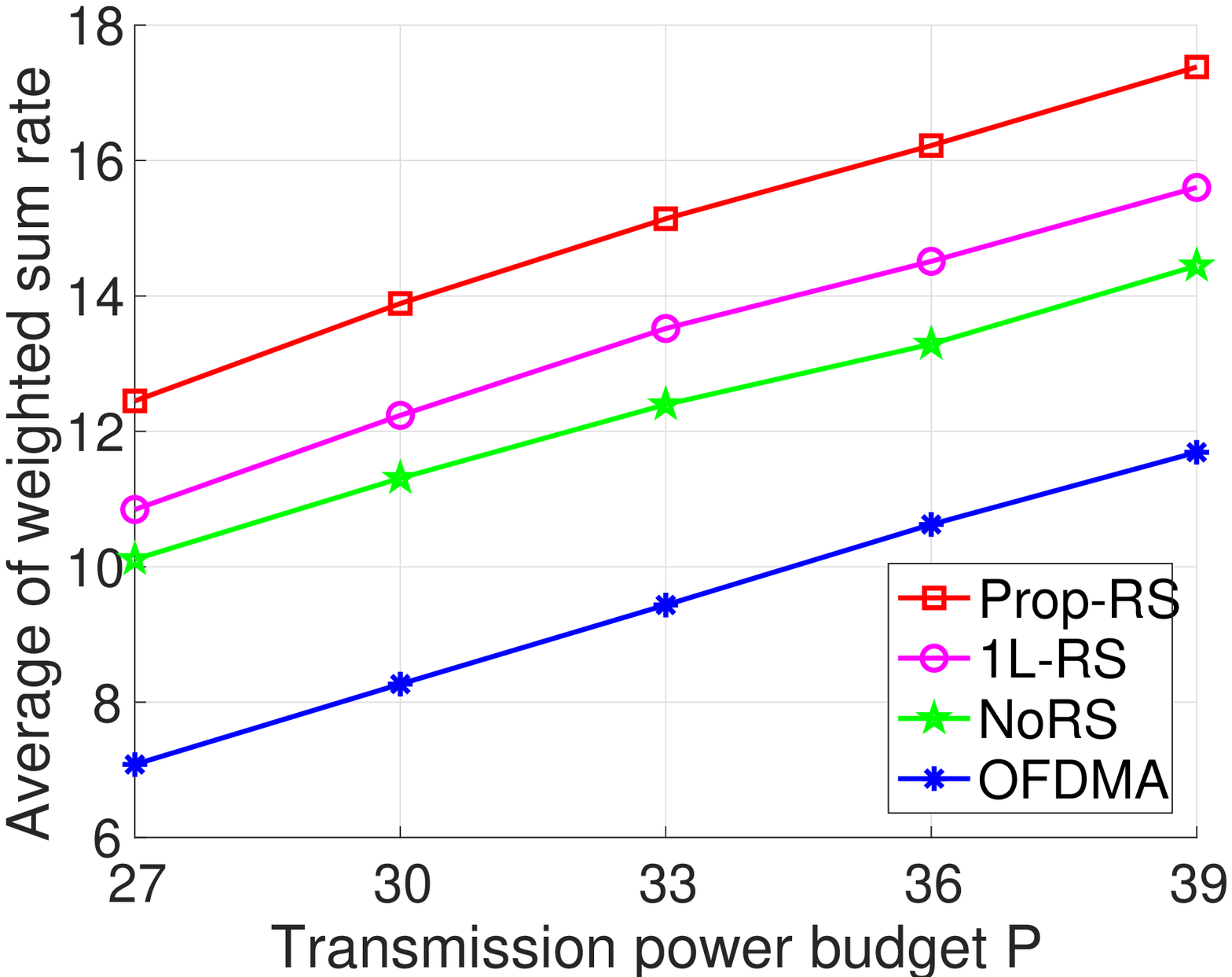}
    \caption{Weighted sum rate versus $P$.}
    \label{fig:label2}
  \end{minipage}
    \begin{minipage}{.45\columnwidth}
    \centering
    \includegraphics[width=\textwidth]{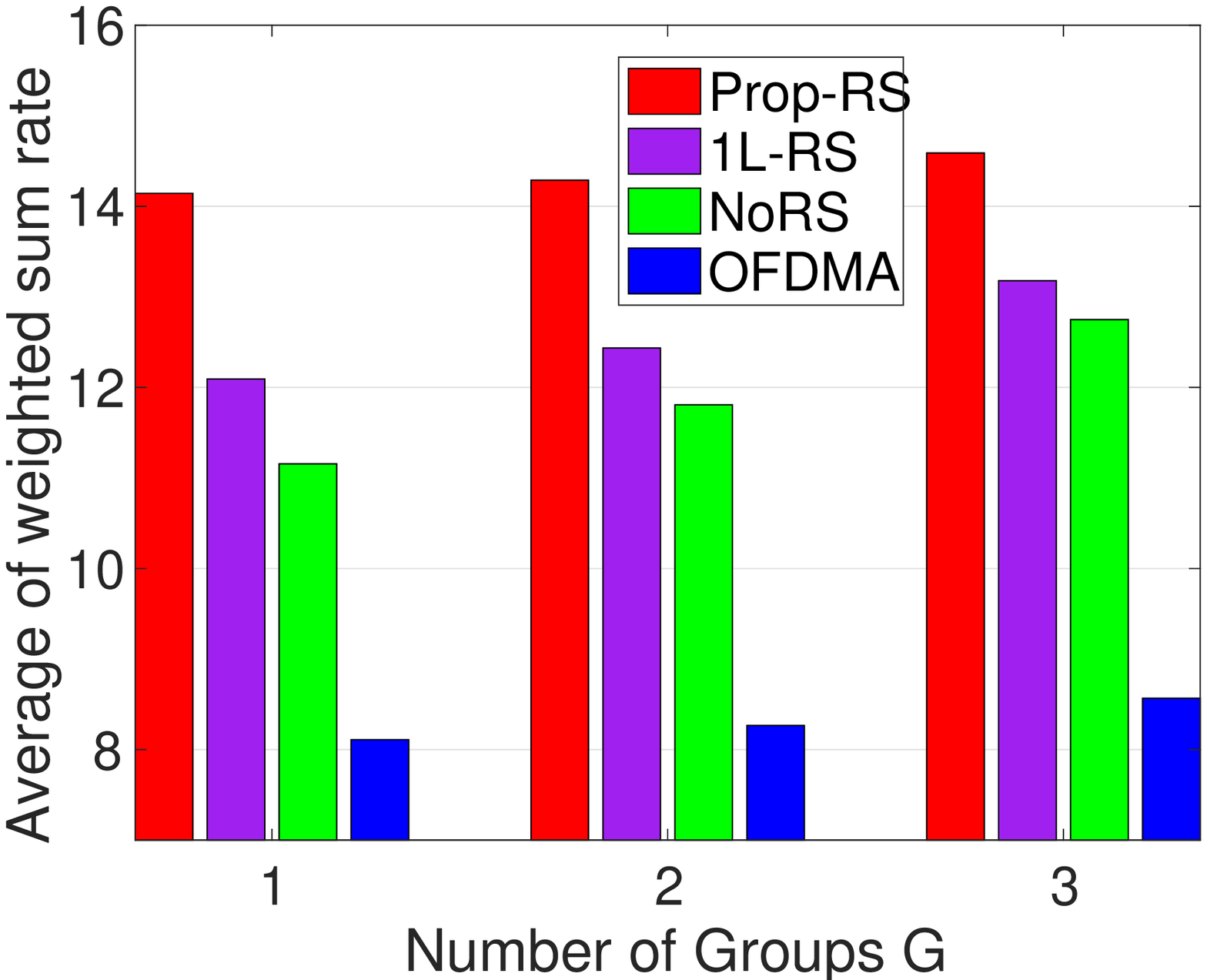}
    \caption{Weighted sum rate versus $G$.}
    \label{fig:label3}
  \end{minipage}
    \begin{minipage}{.45\columnwidth}
    \centering
    \includegraphics[width=\textwidth]{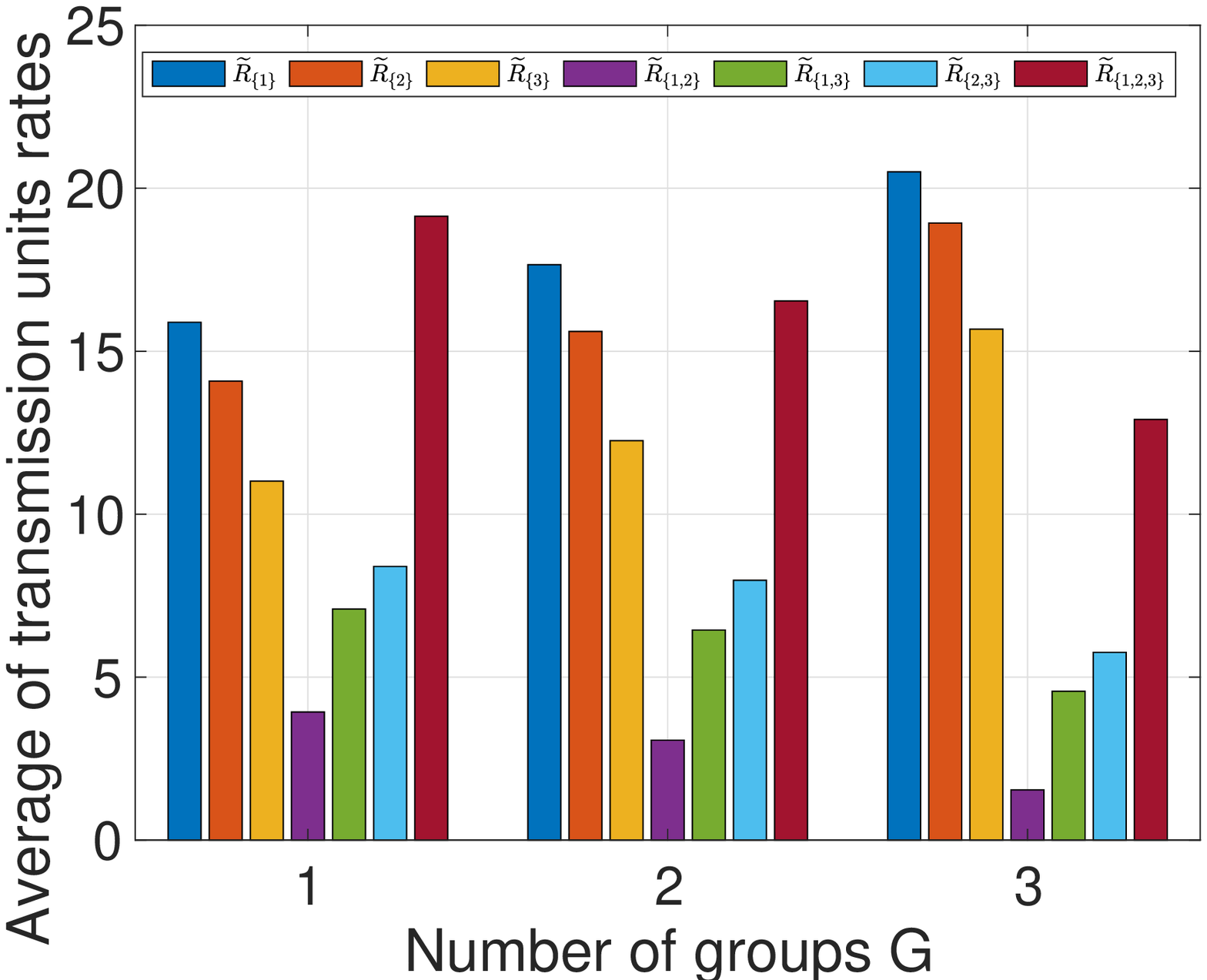}
    \caption{Rates of transmission units of Prop-RS versus $G$.}
    \label{fig:label4}
  \end{minipage}
\end{figure}

Fig.~\ref{fig:label1}, Fig.~\ref{fig:label2}, and Fig.~\ref{fig:label3} illustrate the average of the weighted sum rate versus the number of transmit antennas $M$, the total transmission power budget $P$, and the number of user groups $G$, respectively. From the three figures, we have the following observations.
Firstly, the weighted sum average rate of each scheme increases with $M$, $P$, and $G$. Secondly, Prop-RS outperforms the baseline schemes. The gain of Prop-RS over 1L-RS is because the proposed solution \textcolor{black}{unleashes the full potential} of the flexibility of rate splitting. The gain of Prop-RS over NoRS arises because the cost for NoRS to suppress interference \textcolor{black}{is} high. In contrast, rate splitting together with joint decoding partially decodes interference and partially treats interference as noise. The gain of Prop-RS over OFDMA comes from an effective nonorthogonal transmission design. 
Additionally, Fig.~\ref{fig:label3} shows that \textcolor{black}{the gains of Prop-RS over 1L-RS and NoRS increase as $G$ decreases,} demonstrating the advantage of flexibly dealing with interference in the presence of channel correlation among users.
Fig.~\ref{fig:label4} shows the rates of the transmission units in the proposed solution \textcolor{black}{versus the number of user groups $G$.} We can see that $\widetilde{R}_{\{1\}},\widetilde{R}_{\{2\}},$ and $\widetilde{R}_{\{3\}}$ increase with $G$, whereas $\widetilde{R}_{\{1,2\}},\widetilde{R}_{\{1,3\}},\widetilde{R}_{\{2,3\}},$ and $\widetilde{R}_{\{1,2,3\}}$ decrease with $G$. This is because as channel correlation among the users decreases, it is efficient to decode less interference and treat more interference as noise.

\section{Conclusion}
While applications such as content delivery are responsible for a large and increasing fraction of Internet traffic, general multicast communication will play a central role for future 6G and beyond networks. This paper investigated the \textcolor{black}{optimization} of general rate splitting for general multicast. We adopted linear beamforming at the BS and joint decoding at each user. We maximized the weighted sum rate under the achievable rate region constraints and power constraint. We proposed an iterative algorithm to obtain a KKT point. The proposed optimization framework generalizes the existing ones for rate splitting for unicast, single-group multicast, and multi-group multicast. Numerical results demonstrate notable gains of the proposed solution over existing schemes and reveal the impact of channel correlation among users on the performance of general rate splitting for general multicast. \textcolor{black}{There are still some key aspects that we leave for future investigations. 
One direction is to go beyond linear approaches and investigate nonlinear precoders such as binning. Another interesting perspective is general multicast with partial channel state information at the transmitter side.}


\begin{thebibliography}{10}
\providecommand{\url}[1]{#1}
\csname url@samestyle\endcsname
\providecommand{\newblock}{\relax}
\providecommand{\bibinfo}[2]{#2}
\providecommand{\BIBentrySTDinterwordspacing}{\spaceskip=0pt\relax}
\providecommand{\BIBentryALTinterwordstretchfactor}{4}
\providecommand{\BIBentryALTinterwordspacing}{\spaceskip=\fontdimen2\font plus
\BIBentryALTinterwordstretchfactor\fontdimen3\font minus
  \fontdimen4\font\relax}
\providecommand{\BIBforeignlanguage}[2]{{%
\expandafter\ifx\csname l@#1\endcsname\relax
\typeout{** WARNING: IEEEtran.bst: No hyphenation pattern has been}%
\typeout{** loaded for the language `#1'. Using the pattern for}%
\typeout{** the default language instead.}%
\else
\language=\csname l@#1\endcsname
\fi
#2}}
\providecommand{\BIBdecl}{\relax}
\BIBdecl



\bibitem{TWC21}
C.~Guo, L.~Zhao, Y.~Cui, Z.~Liu, and D.~Ng "Power-efficient transmission of multi-quality tiled 360 VR video in MIMO-OFDMA systems," \emph{IEEE Trans. Wireless Commun.}, vol. 20, no. 8, pp. 5408-5422, Aug. 2021.

\bibitem{TCOM20}
W.~Xu, Y.~Cui, and Z.~Liu, ``Optimal multi-view video transmission in multiuser wireless networks by exploiting natural and view synthesisenabled multicast opportunities," \emph{IEEE Trans. Commun.}, vol. 68, no. 3, pp. 1494-1507, Mar. 2020.


\bibitem{TON17}
Y.~Cui, M.~Médard, E.~Yeh, D.~Leith, F.~Lai, and K.~R.~Duffy, "A linear network code construction for general integer connections based on the constraint satisfaction problem," \emph{IEEE/ACM Trans. Netw.}, vol. 25, no. 6, pp. 3441-3454, Dec. 2017.

\bibitem{ISIT17}
H.~P.~Romero and M.~K.~Varanasi, "Rate splitting, superposition coding and binning for groupcasting over the broadcast channel: A general framework," \emph{arXiv preprint arXiv:2011.04745}, Nov. 2020.

\bibitem{TIT1981}
T.~Han and K.~Kobayashi, ``A new achievable rate region for the interference channel,'' \emph{IEEE Trans. Inf. Theory}, vol. 27, no. 1, pp. 49-60, Jan. 1981.


\bibitem{TIT13_yang}
S.~Yang, M.~Kobayashi, D.~Gesbert, and X.~Yi, ``Degrees of freedom of time correlated MISO broadcast channel with delayed CSIT,'' \emph{IEEE Trans. Inf. Theory}, vol. 59, no. 1, pp. 315-328, Jan. 2013.


\bibitem{arxiv21}
J. Park, J. Choi, N. Lee, W. Shin, and H. V. Poor, ``Rate-splitting multiple access for downlink MIMO: A generalized power iteration approach," \emph{arXiv preprint arXiv:2108.06844}, Aug. 2021.

\bibitem{TSP16}
H.~Joudeh and B.~Clerckx, ``Robust transmission in downlink multiuser
MISO systems: A rate-splitting approach,” \emph{IEEE Trans. Signal Process.}, vol. 64, no. 23, pp. 6227–6242, Dec. 2016.


\bibitem{JSAC21}
Z.~Li, C.~Ye, Y.~Cui, S.~Yang, and S.~Shamai, ``Rate splitting for
  multi-antenna downlink: Precoder design and practical implementation,''
  \emph{IEEE J. Select. Areas Commun.}, vol.~38, no.~8, pp. 1910--1924, Jun. 2020.


\bibitem{TCOM19}
Y.~Mao, B.~Clerckx, and V.~O.~K. Li, ``Rate-splitting for multi-antenna non-orthogonal unicast and multicast transmission: spectral and energy efficiency analysis,"  \emph{IEEE Trans. Commun.}, vol. 67, no. 12, pp. 8754-8770, Dec. 2019.




\bibitem{TVT20}
H.~Chen, D.~Mi, B.~Clerckx, Z.~Chu, J.~Shi, and P.~Xiao, ``Joint power and subcarrier allocation optimization for multigroup multicast systems with rate splitting,'' \emph{IEEE Trans on Veh. Technol.}, vol. 69, no. 2, pp. 2306-2310, Feb. 2020.





\bibitem{TSP06}
N.~D.~Sidiropoulos, T.~N.~Davidson and Z. Luo, ``Transmit beamforming for physical-layer multicasting,'' \emph{IEEE Trans. Signal Process.}, vol.~54, no.~6, pp.~2239-2251, Jun. 2006.




\bibitem{TSP17}
Y.~{Sun}, P.~{Babu}, and D.~P. {Palomar}, ``Majorization-minimization
  algorithms in signal processing, communications, and machine learning,''
  \emph{IEEE Trans. Signal Process.}, vol.~65, no.~3, pp. 794--816, Feb. 2017.




\bibitem{MOR21}
F.~Facchinei, V.~Kungurtsev, L.~Lampariello, and G.~Scutari, ``Ghost
penalties in nonconvex constrained optimization: Diminishing stepsizes
and iteration complexity,'' \emph{Math. Oper. Res.}, vol. 46, no. 2, pp. 595-627, Feb. 2021.


\bibitem{TCOM07}
W.~Choi, A.~Forenza, J.~G.~Andrews, and R.~W.~Heath, ``Opportunistic space-division multiple access with beam selection,'' \emph{IEEE Trans. Commun.}, vol. 55, no. 12, pp. 2371-2380, Dec. 2007.






\end{thebibliography}
\end{document}